\documentclass{PoS}

\def \inte {{\it INTEGRAL\,}}

\def \sw {{\it Swift}}

\def \hcm {\hbox {\ifmmode $ atom cm$^{-2}\else atom cm$^{-2}$\fi}}

\def \apj {ApJ}
\def \apjl {ApJL}

\def \aap {A\&A}
\def \aap {A\&AS}

\def \mnras {MNRAS}

\title{New results with Swift on Supergiant Fast X-ray Transients}

\ShortTitle{New results with Swift on Supergiant Fast X-ray Transients}

\author{ \speaker{L.\ Sidoli},$^a$ P.\ Romano,$^b$  
L.\ Ducci,$^{ac}$ A.\ Paizis,$^a$ S.\ Vercellone,$^b$ 
G.\ Cusumano,$^b$ V.\ La Parola,$^b$
V.\ Mangano,$^b$  
J.A.\ Kennea,$^d$ D.N.\ Burrows,$^d$
H.A.\ Krimm,$^{efg}$  N.\ Gehrels,$^g$ 
V.~Sguera,$^{hi}$  A.\ Bazzano$^i$  \\
\llap{$^a$}INAF, Istituto di Astrofisica Spaziale e Fisica Cosmica, \\
         Via E.\ Bassini 15,   I-20133 Milano,  Italy\\
\llap{$^b$}INAF, Istituto di Astrofisica Spaziale e Fisica Cosmica, \\
         Via U.\ La Malfa 153, I-90146 Palermo, Italy\\
\llap{$^c$}Dipartimento di Fisica e Matematica, Universit\`a degli Studi dell'Insubria, \\
Via Valleggio 11, I-22100 Como, Italy \\
\llap{$^d$}Department of Astronomy and Astrophysics, Pennsylvania State  University, \\
         University Park, PA 16802, USA\\
\llap{$^e$}CRESST/Goddard Space Flight Center, Greenbelt, MD, USA\\
\llap{$^f$}Universities Space Research Association, Columbia, MD, USA\\
\llap{$^g$}NASA/Goddard Space Flight Center, Greenbelt, MD 20771, USA\\
\llap{$^h$}INAF, Istituto di Astrofisica Spaziale e Fisica Cosmica, \\
           Via Gobetti 101, I-40129 Bologna, Italy\\
\llap{$^i$}INAF, Istituto di Astrofisica Spaziale e Fisica Cosmica, \\ 
           Via Fosso del Cavaliere 100, I-00133, Roma, Italy\\
E-mail: \email{sidoli@iasf-milano.inaf.it}
}

\abstract{We report here on the most recent results obtained  on a new class 
of High Mass X--ray Binaries, the Supergiant Fast X--ray Transients.
Since October 2007, we have been performing a monitoring campaign with {\it Swift} of four SFXTs 
(IGR~J17544--2916, XTE~J1739--302, IGR~J16479--4514 and the X--ray pulsar AX~J1841.0-0536)
for about 1-2~ks, 2--3 times per week,
allowing
us to derive the previously unknown long term properties of this new class of sources 
(their duty cycles, spectral properties in outbursts and out-of-outbursts,
temporal behaviour). 
We also report here on additional $Swift$ observations of two SFXTs which are not part of the monitoring: 
IGR~J18483--0311 (observed with $Swift$/XRT during a whole orbital cycle) 
and SAX~J1818.6--1703 (observed for the
first time simultaneously in the energy range 0.3--100 keV during a bright flare). }

\FullConference{The Extreme sky: Sampling the Universe above 10 keV\\
		 October 13-17 2009\\
		 Otranto (Lecce) Italy}

\begin{document}

\section{Supergiant Fast X--ray Transients}

Supergiant Fast X-ray Transients (SFXTs) are transient X--ray 
sources in binary systems  
composed of a compact object and a blue supergiant companion. Although some of them
were discovered before the \inte\ satellite launch in 2002, this new
 class of High Mass X--ray Binaries (HMXBs)
was recognized only after several new peculiar X--ray transients had 
been found because of the Galactic plane survey performed by \inte\  
(\cite{Sguera2005}, \cite{Negueruela2006}).
The members of this new class of sources display 
apparently short outbursts (as observed with \inte), 
characterized by a few hour duration flares and by a high dynamic range 
(1,\,000--100,\,000) between the quiescence (10$^{32}$~erg~s$^{-1}$) and 
the flare peak (10$^{36}$--10$^{37}$~erg~s$^{-1}$). 

%

The spectra are very similar to those of accreting pulsars, displaying a flat power law spectrum below
10~keV, and a high energy cutoff in the range 10--30 keV (e.g. \cite{Smith1998:17391-3021, SidoliPM2006}).
This spectral similarity suggests that all SFXTs host a neutron star as compact object, although only in four sources 
pulsations have been discovered:  
AX~J1841.0$-$0536 ($P_{\rm spin}\sim4.7$\,s, \cite{Bamba2001}), 
IGR~J11215--5952 ($P_{\rm spin}\sim187$\,s, \cite{Swank2007}), 
IGR~J16465--4507  ($P_{\rm spin}\sim228$\,s, \cite{Lutovinov2005}) and 
IGR~J18483--0311 ($P_{\rm spin}\sim21$\,s, \cite{Sguera2007}).
The distribution of their orbital periods (measured in 5 SFXTs) ranges from 3.3~days (IGR~J16479--4514, \cite{Jain2009:16479}) 
to 165~days (in IGR~J11215--5952, \cite{SidoliPM2006, Sidoli2007, Romano2007, Romano2009:11215_2008}).
Periodically recurrent outbursts have been observed in two sources,  IGR~J11215--5952 \cite{SidoliPM2006} 
and IGR~J18483--0311 (\cite{Levine2006:igr18483, Sguera2007}), 
indicative of  outbursts triggered near the periastron
passage in a highly eccentric orbit.

The physical mechanism responsible for their peculiar transient X--ray emission is still an open issue (see \cite{Sidoli2009:cospar} for a review of the different possibilities).
The main proposed mechanisms are related to the structure of the supergiant wind, accreting onto the compact object 
(\cite{zand2005, Walter2007, Sidoli2007, Negueruela2008, Ducci2009}) 
and/or to the neutron star magnetic field (suggested to be magnetar-like) and its (long) spin period (\cite{Grebenev2007, Bozzo2008}).
These different scenarii need a comparison with
the SFXTs observative properties, which were, on the other hand, 
largely unknown before our monitoring campaign with \sw/XRT, which started in October 2007.
The goal of this campaign was to address several crucial issues,
in particular the source status outside
the bright flaring activity: are SFXTs in quiescence (no accretion) when 
they are not in bright flaring activity? or are they still
in accretion, but at a much lower level, too low to be detected by \inte? 
which is the duty cycle of their transient X-ray emission?

The rapid flaring variability together with the association with a massive companion 
is indicative of the fact that SFXTs are wind accretors, similar to the classical 
persistent HMXBs with supergiant companions (like Vela X--1) of the same spectral type.
One of  the main open issues is the link between these two subclasses of massive binaries. 
Both kind of binaries are composed by a compact object (likely a neutron star) and by an OB 
supergiant companion with similar spectral type.
For this reason, it was proposed that different orbital parameters 
(\cite{Negueruela2008}) could probably separate the two subclasses:
circular and narrow orbits in persistent HMXBs, while wide and eccentric orbits in SFXTs.
On the other hand, the recent discovery of the orbital period of 3.3~days in the 
SFXT IGRJ16479--4514  \cite{Jain2009:16479}, is puzzling, because it implies a narrower 
orbit than that shown by several persistent HMXBs. 
This seems to suggest that at least in a few members of the class, the orbit is not 
the main parameter which separates the transient from the persistent behaviour.

Before our monitoring, 
the broad band spectra (0.3--100 keV) were obtained only from 
not simultaneous observations along the whole energy range, 
thus the suggested spectral similarity
between SFXTs and persistently accreting X--ray pulsars 
needed to be tested by simultaneous wide band observations.
The duration of the outburst phase was unknown, except in the case of the periodic SFXT 
IGR~J11215--5952 \cite{Romano2007}, where our monitoring with \sw\ demonstrated that the outburst 
lasts a few days, instead of only a few hours (as previously observed 
by \inte\ during the brightest flares).
Other crucial properties, 
like the distribution of the neutron star magnetic fields, the spin and the orbital periods, 
were completely unknow, as well.

\section{$Swift$ contribution}

We performed a 
long-term monitoring campaign with \sw/XRT of a sample of four SFXTs,
XTE~J1739--302, IGR~J17544--2619 (the two prototypes of the class),
IGR~J16479--4514 and the X--ray pulsar AX~J1841.0$-$0536, in order to try to address all these important and open issues.
The campaign strategy consists of 2 or 3 XRT pointings per source 
per week (about $\sim$1~ks each) to frequently monitor the source status. 
Given the structure of the observing plan, this monitoring can be considered as a 
casual sampling of the source light curves at a resolution of about $\sim 4$\,days. 
We were mainly interested in the monitoring of the long-term properties, 
to get a census of all the outbursts (even the fainter ones, not triggering the \sw/BAT), 
to monitor the onset of each new outburst and to follow the whole outburst duration with
more frequent subsequent observations (in this respect the \sw\ flexibility is a crucial property),
and to get truly simultaneous wide band spectra during bright flaring activity.
Our monitoring campaign  (which is still on-going) has completely changed the previous view of this transient sources
(\cite{Sidoli2008:sfxts_paperI,
Romano2008:sfxts_paperII,
Sidoli2009:sfxts_paperIII,
Sidoli2009:sfxts_paperIV, Romano2009:sfxts_paperV}).

\sw\ observations have demonstrated that SFXTs spend most 
of their life still accreting matter even outside bright flaring 
activity, with an intermediate level of X--ray emission at 
10$^{33}$--10$^{34}$~erg~s$^{-1}$, 
large flux variability (at least one order of magnitude) and an absorbed 
power law spectrum below 10 keV (photon index of 1--2, or hot black body temperatures of 1--2~keV).
Besides the bright outbursts (detected also with BAT) and the intermediate level of X--ray
emission, several 3$\sigma$ upper limits were also measured, either because the source was faint or due to 
a very short exposure time because of the interruption by a gamma-ray burst. 
In order to get an as uniform as possible subsample 
for the ``non-detections'' state, we excluded all observations with a net exposure below 900\,s.
An exposure of 900\,s corresponds to 2--10\,keV flux limits 
of $\sim$1--3$\times 10^{-12}$ erg cm$^{-2}$ s$^{-1}$ (3$\sigma$), 
depending on the source (assuming the best fit absorbed power law model 
for the intermediate state of each source). 
Then, a {\it duty cycle of inactivity} (IDC) was defined as  
the time fraction each source spends undetected down to a flux limit of 
1--3$\times10^{-12}$ erg cm$^{-2}$ s$^{-1}$ (which means an upper limit to the time spent in quiescence).  
The IDCs we obtained during the campaign with $Swift$/XRT are the following: 
17~\% (IGR~J16479--4514), 28~\% (AX~J1841.0--0536), 39~\% (XTE~J1739--302) 
and 55~\% (IGR~J17544--2619).
In the eclipsing SFXT IGR~J16479$-$4514 a main contribution to the 
IDC comes from the X--ray eclipses, 
hence the above 17\,\% is in fact an upper limit to the true quiescent time.
To summarize, the quiescence in these transients is a rarer state 
\cite{Romano2009:sfxts_paperV} than what previously thought 
based only on \inte\ observations.  
The lowest luminosity level we could observe with $Swift$/XRT, 
obtained accumulating all data for which no detections were obtained as single 
exposures \cite{Romano2009:sfxts_paperV}, 
is reached in 
XTE~J1739--302 (6$\times$10$^{32}$~erg~s$^{-1}$, 2--10 keV) and 
in IGR~J17544$-$2619 (3$\times$10$^{32}$~erg~s$^{-1}$).
 
We observed the broad band simultaneous spectra (XRT together with BAT) during 
8 bright flares observed from 3 of the 4 monitored sources. 
The best fits could be obtained with Comptonized models ({\sc compTT} or {\sc bmc} in {\sc xspec}) 
or with an absorbed flat power law model with
high energy cutoff around 10--30 keV 
(see \cite{Romano2008:sfxts_paperII, Sidoli2009:sfxts_paperIII, Sidoli2009:sfxts_paperIV}). 
This spectral shape, and in particular the spectral cutoff, is 
compatible with a neutron star magnetic field
of $\sim$10$^{12}$~G \cite{Coburn2002}, although no cyclotron lines 
have been detected yet in these four sources.
We found evidence for variable absorbing column densities, both in the same 
source (XTE~J1739--302) and different outbursts and within the same outburst, 
indicative of dense clouds of matter composing the supergiant wind passing towards the line of sight.

Another crucial finding of our monitoring campaign is that the SFXTs bright and 
short flares (a few hour long) are 
part of a longer outburst phase lasting days \cite{Sidoli2009:sfxts_paperIII}, 
as already found in the periodic SFXT IGR~J11215--5952 \cite{Romano2007}.

The first optical/UV observations performed with UVOT 
simultaneously to our \sw/XRT monitoring of the SFXTs revealed a possible hint 
of an UV flaring activity simultaneously to the
X--ray bright flares in XTE~J1739$-$302.
However this findings needs confirmation because it is a less 
than 3 $\sigma$ result \cite{Romano2009:sfxts_paperV}.

\section{SAX~J1818.6--1703}

SAX~J1818.6--1703 is a Supergiant Fast X-ray Transient 
associated with a B0.5Iab type companion located at 2.1~kpc
(\cite{Torrejon2009}).
It triggered the $Swift$/BAT on 2009 May 5 at 
14:03:27 UT, and after the BAT trigger, 
the whole outburst evolution and the decline phase could be
 monitored with XRT.
The time resolved spectroscopy with XRT did not result in 
variability of the spectral parameters
in the 1--10~keV range, within the uncertainties.
No variability in the absorbing column density could be 
detected along the outburst, as well.

Simultaneous BAT and XRT spectra, between 138 and 937\,s 
since the BAT trigger, allowed us to obtain the first 
broad band X--ray spectrum of this source during an outburst, 
with a joint fit in the 1--10\,keV and 
14--150\,keV energy bands for XRT and BAT, respectively. 
This wide band X--ray spectrum, highly absorbed 
($N_{\rm H}$$\sim$5--7$\times$10$^{22}$~cm$^{-2}$), is well 
deconvolved with models like power laws with high energy cutoffs 
({\sc cutoffpl} in {\sc xspec}), 
or Comptonization models [{\sc Comptt}  \cite{Titarchuk1994} or the  
{\sc bmc} model  \cite{TMK1996} in {\sc xspec}].
Adopting the {\sc Comptt} model, the properties of the Comptonizing
corona could be constrained well with an 
electron temperature $kT_{e}$$\sim$5--7~keV 
and an optical depth $\tau$$\sim$10 (in a spherical geometry).
These properties are reminiscent of the X--ray spectral shape
of the prototype of the SFXT class, XTE~J1739--302.

\section{IGR~J18483--0311}

The SFXT IGR~J18483$-$0311 is an X--ray pulsar 
($\sim$21 s, \cite{Sguera2007}) and is the second member of the class
where periodically recurring outbursts have been discovered with a period of 
 $18.55\pm{0.03}$~days (\cite{Levine2006:igr18483, Sguera2007}), 
which is very likely the orbital period of the system.
It is associated with a blue supergiant (B0.5Ia) at 
a distance of 3--4 kpc \cite{Rahoui2008:18483}.

We monitored with \sw\ an entire orbital phase (28 days for a 
total on-source exposure of $\sim44$\,ks), 
starting on 2009 June 11 with 2 ks per day (\cite{Romano2009:igr18483}).
The XRT observations shows a highly modulated light curve
with two maxima, separated by a time
interval consistent with the orbital period 
of $\sim$18.5~days. 
A lower limit of 1200 to the dynamical range can 
be obtained from the observed light curve.
The different duration of the two outburst peaks
 monitored with \sw\ 
is likely  resulting from both a different sampling and a 
high intrinsic X--ray variability.
The second peak has a duration of several days, as
 previously observed by INTEGRAL \cite{Sguera2007}.

We interpret the modulation of the light curve with the orbital 
phase as wind accretion along a highly eccentric orbit.
These observations allow to constrain the different
mechanisms proposed to explain the nature of the new class of SFXTs.
Applying the new clumpy wind model for blue supergiants developed
 by \cite{Ducci2009} to the evolution of the observed X--ray 
light curve, we found that, assuming an eccentricity of $e=0.4$,
the X--ray emission from this source can be explained in terms of 
the accretion from a spherically 
simmetric clumpy wind, composed by clumps with masses ranging from 
$10^{18}$~g to 5$ \times 10^{21}$~g (see Ducci et al., these
proceedings for details of the model).

\end{document}